# Physics-Informed Neural Networks for Solving Multiscale Mode-Resolved Phonon Boltzmann Transport Equation

Ruiyang Li[1], Eungkyu Lee[2]*, and Tengfei Luo[1,3,4]*

[1] Department of Aerospace and Mechanical Engineering, University of Notre Dame, Notre Dame, Indiana 46556, USA

[2] Department of Electronic Engineering, Kyung Hee University, Yongin-si, Gyeonggi-do 17104, South Korea

[3] Department of Chemical and Biomolecular Engineering, University of Notre Dame, Notre Dame, Indiana 46556, USA

[4] Center for Sustainable Energy of Notre Dame (ND Energy), University of Notre Dame, Notre Dame, Indiana 46556, USA

* Corresponding author. Email: eleest@khu.ac.kr (E. Lee); tluo@nd.edu (T. Luo)

# Abstract

Boltzmann transport equation (BTE) is an ideal tool to describe the multiscale phonon transport phenomena, which are critical to applications like microelectronics cooling. Numerically solving phonon BTE is extremely computationally challenging due to the high dimensionality of such problems, especially when mode-resolved properties are considered. In this work, we demonstrate the use of physics-informed neural networks (PINNs) to efficiently solve phonon BTE for multiscale thermal transport problems with the consideration of phonon dispersion and polarization. In particular, a PINN framework is devised to predict the phonon energy distribution by minimizing the residuals of governing equations and boundary conditions, without the need for any labeled training data. Moreover, geometric parameters, such as the characteristic length scale, are included as a part of the input to PINN, which enables learning BTE solutions in a parametric setting. The effectiveness of the present scheme is demonstrated by solving a number of phonon transport problems in different spatial dimensions (from 1D to 3D). Compared to existing numerical BTE solvers, the proposed method exhibits superiority in efficiency and accuracy, showing great promises for practical applications, such as the thermal design of electronic devices.

# 1. Introduction

Effective thermal management has long been critical to the design and advancement of electronic devices. With the rapid development of micro/nano technology, the characteristic length of such devices becomes comparable or even smaller than the mean free path of phonons, which are the dominant thermal energy carriers in semiconductors. In these conditions, thermal transport is not purely diffusive, and thus it cannot be accurately described by the conventional Fourier's law[1,2,3]. To describe the multiscale thermal transport process from ballistic to diffusive regions, phonon Boltzmann transport equation (BTE) has been widely used when phase coherence effects are not important[4,5]. Since atomistic level simulation techniques, such as molecular dynamics and first-principles calculations, are computationally impractical for device-level thermal analysis, mesoscopic approaches based on the solution of phonon BTE provide a balance between computational complexity and accuracy.

However, solving the phonon BTE is a challenging task because of its high dimensionality. The BTE for phonons is a nonlinear integro-differential equation with seven independent variables: phonon frequency, polarization, time, three spatial coordinates (x, y, z directions), and two directional angles (polar angle and azimuthal angle). To make the solution to the phonon BTE more computationally tractable, simplifications such as assuming all phonon modes with the same properties (i.e., gray model) are commonly used[6,7,8,9]. However, such treatment can lead to significant inaccuracy in the solutions[10,11]. This is because different phonon modes can have a large span of mean free paths, and they behave differently at a given physical length scale. To address this issue, many numerical BTE solvers have been proposed, such as the Monte Carlo (MC) method[12,13,14,15], the lattice Boltzmann method[16,17], and deterministic discretization-based methods[18,19,20,21]. While the MC method has achieved great success in accounting for dispersion, polarization and various scattering mechanisms, it suffers from stochastic statistical uncertainty and becomes prohibitively inefficient in or near the diffusive regime due to the restrictions on time step and grid size[13, 14]. Improvements in MC methods have recently been made for solving 3D systems based on variation reduction techniques[15, 22]. Different lattice Boltzmann methods have also been used in recent years, but there are still unphysical predictions reported for cases in the ballistic regime[23, 24]. The discrete ordinate method (DOM), which solves the BTE directly using finite element or finite difference methods, discretizes the angular domain into small solid angles to capture the highly non-equilibrium phonon distribution. However, DOM and its variants usually show slow convergence near the diffusive limit and require large memory for solutions[25, 26]. To solve the phonon BTE consistently over a wide length scale range, characterized by the Knudsen number ($Kn$, the ratio of the phonon mean free path to the characteristic length of the system), the hybrid ballistic-diffusive method[27, 28] introduces a cutoff Knudsen number to separate frequency intervals for which the DOM or the Fourier's Law should be used. Although it can be applied to problems at different length scales, the proper selection of the cutoff Knudsen number remains a question and the improved accuracy comes at the cost of increased computational time. Recently, the discrete unified gas kinetics scheme (DUGKS)[29, 30] and implicit kinetic scheme (IKS)[31] have been shown to work efficiently and accurately for low-dimensional thermal transport problems, but they have not been used for solving phonon BTE in 3D geometries.

It can be seen that all the aforementioned numerical schemes have some deficiencies, and few of them can be universally accurate for a wide range of length scales. Moreover, most methods have been only demonstrated for simple 1D or 2D systems, because they usually require massive computational resources for 3D geometries. For instance, while deterministic numerical methods such as IKS can solve 1D and 2D problems accurately within a few minutes[31], it can take several hours to simulate a 3D system for a single time step on a parallel machine with 400 processors[21]. Therefore, there is a pressing need to develop a numerical method that is accurate, efficient, easy to use and able to deal with high dimensional cases.

Recently, machine learning (ML) methods have been leveraged to help predict thermal properties at the atomistic scale (1-10 nm), such as the development of ML potentials for molecular dynamics simulations[32, 33, 34, 35]. However, these efforts are not sufficient to model thermal transport phenomena in the meso- and micro-scales (e.g., 10 nm - 100 μm), which are more relevant to device applications. In the meantime, as deep neural networks (DNNs) possess the capability of accurately approximating any continuous function[36, 37, 38], they can be used to approximate solutions of partial differential equations (PDEs). Leveraging this property, physics-informed neural networks (PINNs) have emerged recently[39, 40, 41, 42], which incorporate the PDE residuals into the cost function and train the solutions using fully-connected DNNs. When the governing PDEs are known, the solutions can be learned in a physics-constrained manner without the need for any labeled training data, which is known as a data-free ML method. PINNs have been successfully employed to simulate forward and inverse problems for a variety of PDEs[43, 44, 45, 46, 47, 48]. Compared to the conventional mesh-based methods (e.g., finite element method), PINNs avoid discretizing the PDE by taking advantage of the automatic differentiation of DNNs[49]. It has also been shown that in many problems DNN as a universal function approximator can overcome the curse of dimensionality, since the number of parameters of a DNN grows at most polynomially in both the reciprocal of the prescribed accuracy $\varepsilon$ and the dimension $d$[50, 51]. In addition, PINNs have been shown capable of learning solutions of PDEs in parameterized spaces (e.g., variable initial/boundary conditions, geometry, equation parameters)[47, 52]. This feature is especially valuable for optimization applications since a numerical procedure for solving PDE has to be performed each time any parameter is changed using conventional solvers – a problem also applicable to BTE solvers. In multiscale thermal transport problems, one such variable parameter could be the characteristic length. However, research efforts for solving PDEs with PINNs have been mainly in the fluid dynamics field aiming to solve Navier-Stokes equations, which usually have three input parameters. The mode-resolved phonon BTE, which has at least seven variables, can potentially benefit even more from the PINN scheme, but it has never been studied.

In this work, we demonstrate the use of PINNs to efficiently solve mode-resolved phonon BTE for multiscale thermal transport problems. In particular, a PINN framework is devised to predict the phonon energy distribution by minimizing the residuals of governing equations and boundary conditions, without the need for any labeled training data. Moreover, geometric parameters, such as the characteristic length scale, are included as a part of the input to PINN for learning BTE solutions in a parametric setting. This enables our method to handle structures over a wide range of length scales after a single training. The effectiveness of the present scheme is demonstrated by solving a number of phonon transport problems in different spatial dimensions (from 1D to 3D). Compared to conventional numerical BTE solvers, the proposed method exhibits superiority in efficiency and accuracy, showing great promises for practical applications, such as the thermal design of electronic devices.

## 2. Results

**PINNs for stationary phonon BTE**

It can be shown that for a system without internal heat source at steady state, the multiscale thermal transport problem can be described by the BTE with the single mode relaxation time approximation (see the Methods section for derivation):

$$\mathcal{R}(e(\boldsymbol{x},\boldsymbol{s},k,p,\boldsymbol{\mu})) = 0 := \begin{cases} \boldsymbol{v}\cdot\nabla e - \dfrac{e^{eq} - e}{\tau} = \boldsymbol{v}\cdot\nabla e + \dfrac{e^{neq}}{\tau} = 0 \\ \nabla\cdot\boldsymbol{q} = \nabla\cdot\sum_{p}\int_0^{\omega_{max,p}}\int_{4\pi}\boldsymbol{v}e d\Omega\, d\omega = 0, \end{cases} \quad \boldsymbol{x},\boldsymbol{s},k,p\in\Gamma, \boldsymbol{\mu}\in\mathbb{R}^d, \quad (1)$$

where $e$ represents phonon energy distribution function of variables (spatial coordinates $\boldsymbol{x}$, directions $\boldsymbol{s}$, wave number $k$, and phonon polarization $p$) in domain $\Gamma$ and parameters $\boldsymbol{\mu}$, $\tau$ is the relaxation time of phonon with $k$ and $p$, and $e^{neq} = e - e^{eq}$ is the non-equilibrium part of the distribution function. Here, parameters $\boldsymbol{\mu}$ include the characteristic length of the simulation system and phonon mean free path. The heat flux $\boldsymbol{q}$ is calculated by integrating $e$ over the solid angle space ($\Omega$) and frequency space ($\omega$, $p$). The solutions of $e$ can be determined when suitable boundary conditions are prescribed,

$$\mathcal{B}_i(\boldsymbol{x},\boldsymbol{s},k,p,e,\boldsymbol{\mu}) = 0, \quad \boldsymbol{x},\boldsymbol{s},k,p\in\Gamma_b, \boldsymbol{\mu}\in\mathbb{R}^d, \quad (2)$$

where $\mathcal{B}_i$ are operators that define various boundary conditions, and $\Gamma_b$ denotes the boundary region. In the Methods section, we show three typical boundary conditions encountered in phonon BTE, including the isothermal boundary, the diffusely reflecting boundary, and the periodic boundary.

For a given set of parameters $\boldsymbol{\mu}$, $e$ can be obtained by conventional numerical BTE solvers, which usually involve mesh generation and iteratively solving large linear/nonlinear systems. This process can be extremely computationally costly for mode-resolved phonon BTE given the large number of input variables. In addition, solving problems with variable geometries is further challenging with these methods as any change of the geometry requires regeneration of the mesh and performing the calculations over again. Thus, we have not seen many studies adopting mode-resolved models for 3D solutions.

To enable fast predictions of steady-state multiscale thermal transport, we develop a PINN model to approximate the solutions of phonon BTE (Eq. (1)) with the length scale of simulation domain as an additional parameter included in parameters $\boldsymbol{\mu}$. It is the inclusion of such physical quantities as input variables that makes our PINN scheme capable of solving BTE in the parametric setting. This PINN framework, as shown in Fig. 1, is expected to offer a rapid online evaluation of the phonon transport with any given characteristic dimension of the target structure after offline pre-training. The input layer is composed of $\boldsymbol{x}$, $\boldsymbol{s}$, $k$, $p$, mean free path $\lambda = |\boldsymbol{v}|\tau$, and length scale $L$. Apart from the length scale, the mean free path is also included as an input parameter because it impacts Eq. (1) for each phonon mode through relaxation time. Given that the magnitudes of $e^{eq}$ and $e^{neq}$ can be very different when the frequency-dependent Knudsen number $Kn = \lambda/L$ is extremely small (e.g., $Kn \sim O(10^{-4})$), two fully-connected DNNs are used to approximate the equilibrium and non-equilibrium distribution functions, respectively. Each sub-network maps the inputs to a target output, through several layers of neurons consisting of either affine linear transformations or scalar non-linear activation functions. The DNNs are trained by minimizing a weighted sum of residuals of Eqs. (1) and (2) as follows:

$$\mathcal{L}(\boldsymbol{W}, \boldsymbol{b}) = \left\| \boldsymbol{v} \cdot \nabla e - \frac{e^{eq} - e}{\tau} \right\|^2 + \|\nabla \cdot \boldsymbol{q}\|^2 + \sum_i \|\mathcal{B}_i\|^2, \qquad (3)$$

where $\boldsymbol{W}$ and $\boldsymbol{b}$ refer to the weights and biases of the entire network, and $\|\cdot\|$ is $L_2$ norm. The boundary conditions are formulated as additional penalty terms into the loss function and imposed in a soft manner, i.e., formulated as part of the loss function instead of directly imposed in the solution. An optimal set of network parameters is identified by minimizing this composite loss function,

$$\boldsymbol{W}^*, \boldsymbol{b}^* = \arg\min_{\boldsymbol{W},\boldsymbol{b}} \mathcal{L}(\boldsymbol{W}, \boldsymbol{b}). \quad (4)$$

It is noted that several first derivatives of $e$ with respect to spatial coordinates $\boldsymbol{x}$ are required for constructing the PDE residuals. These derivatives can be computed accurately and efficiently by automatic differentiation in the DNN, which uses the chain rule to back-propagate derivatives from the output layer to the inputs. This approach yields a smooth representation of the solutions that can be evaluated with any input in a certain domain of interest. The above optimization problem is solved by stochastic gradient descent algorithms. The specific DNN architecture and training details are included in the Methods section.

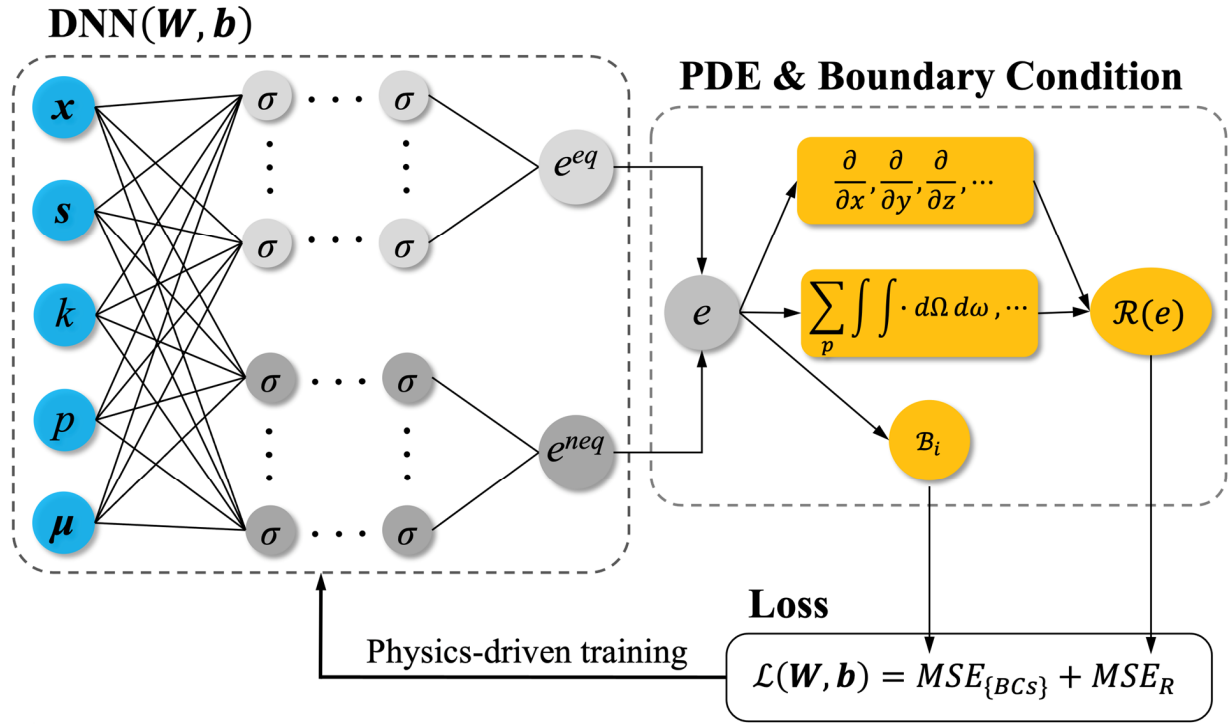

**Fig. 1 Schematic of PINN framework for solving stationary mode-resolved phonon BTE.** Two DNNs are employed to approximate the equilibrium ($e^{eq}$) and non-equilibrium ($e^{neq}$) parts of the phonon energy distribution, respectively. Inputs include spatial vector $\boldsymbol{x}$, directional unit vector $\boldsymbol{s} =$

$(cos\theta, sin\theta cos\varphi, sin\theta sin\varphi)$ ($\theta$ is the polar angle and $\varphi$ is the azimuthal angle), wave number $k$ and polarization $p$. $\boldsymbol{\mu}$ represents additional parameters, which include modal mean free path $\lambda$ and system characteristic length $L$ in this study. $\sigma$ represents the activation function, which is set to be the Swish activation function in this work. The loss function of PINN contains residuals of the PDEs and boundary conditions on a sample of training points in the domain of interest. The hyperparameters in the DNNs are initialized by He's method[53] and learned by minimizing the total loss.

**Model Systems**

To evaluate the performance of the proposed PINN scheme, several steady-state thermal transport problems are investigated at different characteristic length scales, including 1D cross-plane, 2D in-plane, 2D square and 3D cuboid heat conduction problems. We use single crystalline silicon as a model material system, which is the most representative semiconductor in electronics, but the proposed PINN scheme is applicable to other materials. The phonon dispersion relation of silicon in the [100] direction is used, and isotropy is assumed[54]. Only one longitudinal acoustic (LA) and two degenerate transverse acoustic (TA) phonon branches are considered because the optical branches contribute little to the thermal transport[14]. We note that including more branches is possible and only involves expanding the input space to the PINN. The phonon dispersion relation and relaxation times are determined based on Refs. 55, 56 (see Supplementary Note 1 for the phonon dispersion and relaxation times considered). As we assume that the silicon simulated is not heavily doped like those in real electronics, electron-phonon interactions are not considered important in this study, but this effect can be easily included by adding its influence in the phonon relaxation times[57]. For each phonon branch, the wave vector space $k \in [0, 2\pi/a]$ is equally discretized into $N_B$ frequency bands, where $a = 5.431$ Å is the lattice constant for silicon. The corresponding bulk thermal conductivity at 300 K for 10 frequency bands is 145.6 W/(m·K), which is close to the literature value[58], and thus we set $N_B = 10$ in all subsequent calculations. In all cases, the reference temperature $T_{ref}$ is set as 300 K. The training and testing information about numerical experiments are summarized in Table 1. For a given length scale $L$, $N_x$ is the number of interior points in the spatial domain, and $N_s$ is the number of solid angles. $N_l$ represents the number of length scale parameters $L$ sampled in a certain range in Table 1. The training and validation losses of all cases are shown in Table 2, where the training loss is defined in Eq. (3) after nondimensionalization and the validation loss is the total loss evaluated in testing with the same settings shown in Table 1.

**Table 1 Training and testing information about numerical experiments.** Walltimes are on a NVIDIA GeForce TITAN Xp GPU. Two models are trained for 3D geometries with different ranges of length scale $L$, as denoted with superscript a and b.

| Case | Training | | | | Testing | | | | $L$ (m) |
|---|---|---|---|---|---|---|---|---|---|
| | $N_x$ | $N_s$ | $N_l$ | Walltime (h) | $N_x$ | $N_s$ | $N_l$ | Walltime (s) | |
| 1D cross-plane | 40 | 16 | 5 | 0.80 | 40 | 64 | 17 | 2.11 | $[10^{-8}, 10^{-4}]$ |
| 2D in-plane | 300 | 144 | 5 | 4.83 | 1600 | 1024 | 17 | 24.72 | $[10^{-8}, 10^{-4}]$ |
| 2D square | 450 | 144 | 4 | 6.62 | 2601 | 576 | 7 | 9.57 | $[10^{-8}, 10^{-5}]$ |
| 3D cuboid[a] | 1200 | 144 | 3 | 15.52 | 132651 | 576 | 5 | 344.54 | $[10^{-8}, 10^{-6}]$ |
| 3D cuboid[b] | 1200 | 64 | 3 | 8.68 | 132651 | 576 | 5 | 351.64 | $[10^{-5}, 10^{-3}]$ |

**Table 2 Training and validation losses of numerical experiments.**

| Case | Training Loss | Validation Loss |
|---|---|---|
| 1D cross-plane | $2.0 \times 10^{-4}$ | $1.4 \times 10^{-3}$ |
| 2D in-plane | $5.6 \times 10^{-4}$ | $4.5 \times 10^{-3}$ |
| 2D square | $2.4 \times 10^{-2}$ | $1.6 \times 10^{-2}$ |
| 3D cuboid[a] | $6.5 \times 10^{-3}$ | $8.8 \times 10^{-3}$ |
| 3D cuboid[b] | $9.8 \times 10^{-3}$ | $4.9 \times 10^{-3}$ |

**1D cross-plane phonon transport**

We first consider the heat conduction across a silicon thin film (see inset in Fig. 2a). Phonon-mediated heat conduction in thin films plays a crucial role in nanoscale devices, and much effort has been devoted to the measurement of cross-plane thermal conductivity of thin films[59, 60]. Computationally, conventional numerical methods have been proved capable of yielding good predictions of cross-plane heat conduction[29, 31, 61]. Given that the lateral dimensions of the film are much larger than the thickness, the heat conduction reduces to an 1D problem governed by an 1D phonon BTE with two isothermal boundary conditions (see the Methods section). The thickness of the film is $L$, and the left (right) boundary is fixed at $T_L = T_{ref} + \Delta T/2$ ($T_R = T_{ref} - \Delta T/2$) with $\Delta T$ set to 1 K.

To train our PINN model, the spatial domain is equally discretized with 40 interior training points, and 16-point Gauss-Legendre quadrature is used for the direction $\mathbf{s}_x = cos\theta$ (Table 1). For the purpose of learning the parametric solutions for structures of varying characteristic dimensions, a vector of log values of thickness $L$ is included as a part of the input ($\boldsymbol{\mu}$ in Fig. 1) to the network. The model is trained on five values of $L$ in the range between 10 nm and 100 μm. After training, the temperature and heat flux can be evaluated at new length scales given the interpolation ability of DNNs.

Figure 2a shows the dimensionless temperature profiles $T^* = (T - T_R)/(T_L - T_R)$ with different thicknesses $L$. The results from IKS[31] are included for comparison. It is observed that the temperatures evaluated by the PINN model agree almost perfectly with those predicted by the deterministic IKS numerical solver. The discrepancy of these two models is less than 0.6 %. For cases with small thicknesses, PINN successfully captures the temperature slips near the boundary because of the non-equilibrium phonon transport. We also compute the dimensionless thermal conductivities ($k_{eff}/k_{bulk}$) at different length scales, where $k_{eff} = qL/\Delta T$ is the effective thermal conductivity if Fourier's law were used, and $k_{bulk} = \frac{1}{3}\Sigma_p \int_0^{\omega_{max,p}} C|\boldsymbol{v}|^2 \tau d\omega$ represents the bulk thermal conductivity in the diffusive limit. As shown in Fig. 2b, good agreement is observed for all cases (error < 1 %). While only trained on five parameter points in characteristic length (i.e., thickness in this case), this model is able to provide accurate predictions of thermal conductivity at unseen input thicknesses.

As for the computational cost, the training time of one PINN scheme in this case is approximately 48 minutes as shown in Table 1, but the online evaluation takes only a few seconds depending on the input size. Although existing numerical solvers can also compute a simple 1D case efficiently, PINN offers an approach to solving multiscale phonon BTE in a parametric setting, where the length scale is incorporated as an input variable and can be set freely in the range of interest in training. This feature can accelerate the calculation significantly.

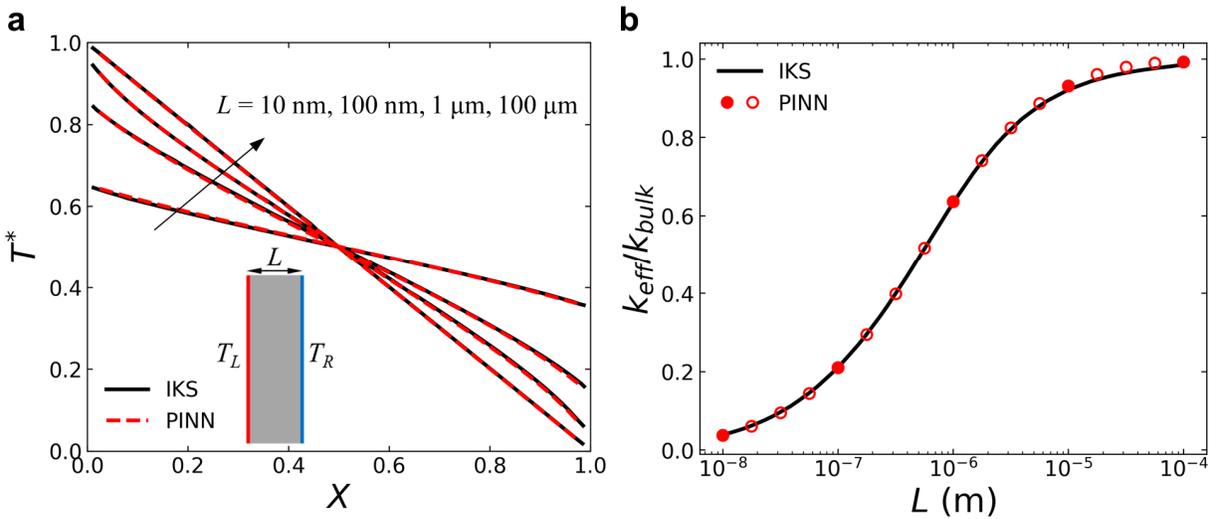

**Fig. 2 Results of 1D cross-plane phonon transport**. a Dimensionless temperature profiles of silicon thin films with different thicknesses ($L$ = 10 nm, 100 nm, 1 μm, 100 μm), where $T^* = (T - T_R)/(T_L - T_R)$ and $X = x/L$. Inset shows the film geometry and boundary conditions. The black solid lines are from the Implicit Kinetic Scheme (IKS) solution to BTE in Ref. 31. **b** Effective thermal conductivity normalized by the bulk thermal conductivity at different film thicknesses. The filled circles represent the parameter points used for training, while the hollow circles are predicted points not included in training.

**2D in-plane phonon transport**

The next case is in-plane thermal transport in two spatial dimensions. We consider the setup depicted in Fig. 3a, where a silicon film is bounded by two adiabatic walls with a distance of $H$ in the y-direction. A uniform temperature gradient is imposed along the x-direction. Diffusely reflecting boundary condition (see the Methods section) is applied for the two adiabatic walls, while the left and right boundaries are treated as periodic boundaries. In the spatial domain, 300 Sobol points for a square-shape regime are used as the interior training points, along with 30 equidistant points used for each boundary (Supplementary Note 2). To avoid the ray effect[62] induced by the conventional $S_N$ quadrature, we discretize $cos\theta$ in $[-1, 1]$ and the azimuthal angle $\varphi$ in $[0, \pi]$ (due to symmetry) into 12 × 12 discrete angles using Gauss-Legendre quadrature. In this 2D problem, the characteristic length $H$ is treated as an input variable parameter, and we train a PINN model to predict the phonon transport at the length scale ranging from 10 nm to 100 μm.

Figure 3b shows the dimensionless x-directional heat flux $q_x^* = q_x(Y)/q_{bulk}$ for different $H$ values, where $q_{bulk} = -k_{bulk} \cdot dT/dx$. The results are consistent with the analytical solutions by the Fuchs-Sondheimer theory[63] over a wide range of length scales, with an prediction error less than 2.4 %. We also compare the effective thermal conductivity $k_{eff} = -(dT/dx)^{-1} \int_0^1 q_x(Y)dY$ with the benchmarks. As shown in Fig. 3c, predictions by PINN agree well with the analytical solutions (error < 1.6 %), and again this method captures the thermal conductivity variation due to the size change. It should be mentioned that all the evaluations are performed on a much denser grid mesh than the one used in training. The training takes about 4.83 hours, while the evaluation costs less than 25 seconds for the total of 17 different $H$ values with a 40 × 40 spatial mesh and a 32 × 32 angular mesh (see Table 1).

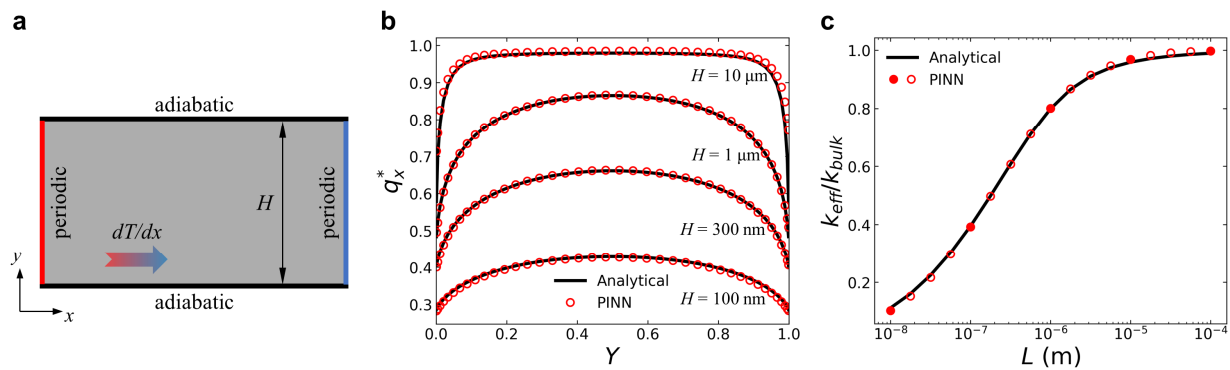

**Fig. 3 Results of 2D in-plane phonon transport**. **a** Schematic of 2D in-plane phonon transport case. **b** Distribution of dimensionless x-directional heat flux along the y-axis in thin films with different thicknesses, where $q_x^* = q_x(Y)/q_{bulk}$ and $Y = y/H$. From bottom to top, the PINN predictions are shown as red circles for $H$ = 100 nm, 300 nm, 1 μm, 10 μm. The black solid lines denote analytical solutions by the Fuchs-Sondheimer theory[63]. **c** Effective thermal conductivity normalized by the bulk thermal conductivity at different length scales. The filled circles represent the parameter points used for training, while the hollow circles are predicted points not included in training.

**2D square phonon transport**

We further consider a case of a 2D square silicon as illustrated in Fig. 4a. The temperature at the top boundary is maintained at $T_h$, while all other boundaries are held at a lower temperature $T_c$. The temperature difference is set as $\Delta T = T_h - T_c = 1\ K$, and the isothermal boundary conditions are imposed on all the boundaries.

In this case, we employ a denser spatial grid mesh for improved accuracy near the top boundary (Supplementary Note 2). Similarly, the training is conducted on only four geometries with length $L$ ranging from 10 nm to 10 μm, and the trained model can be tested at any input length scale in this range. It is noted that the validation loss is lower than the training loss (Table 2). This is because we use a non-uniform spatial mesh with local refinement near the top boundary in training, while in testing we adopt a uniform mesh.

Although there is no available numerical result for direct comparison in this case, we show the temperature profiles along the vertical centerline at $x = 0.5L$ in Fig. 4b, together with the analytical solution to the 2D steady-state Fourier heat equation (i.e., solution in the diffusive limit). It is observed that the result for the 10 μm case is nearly identical to the analytical solution. We note that the average mean free path of phonons in silicon is around 300 nm, which is much shorter than 10 μm. Thus, the 10 μm case is supposed to be close to the diffusive limit. Figures 4c-f show the 2D dimensionless temperature contour $T^* = (T - T_c)/\Delta T$ at different length scales. We see obvious temperature slips near the hot boundary for the cases with smaller $L$, and as $L$ increases the slip decreases.

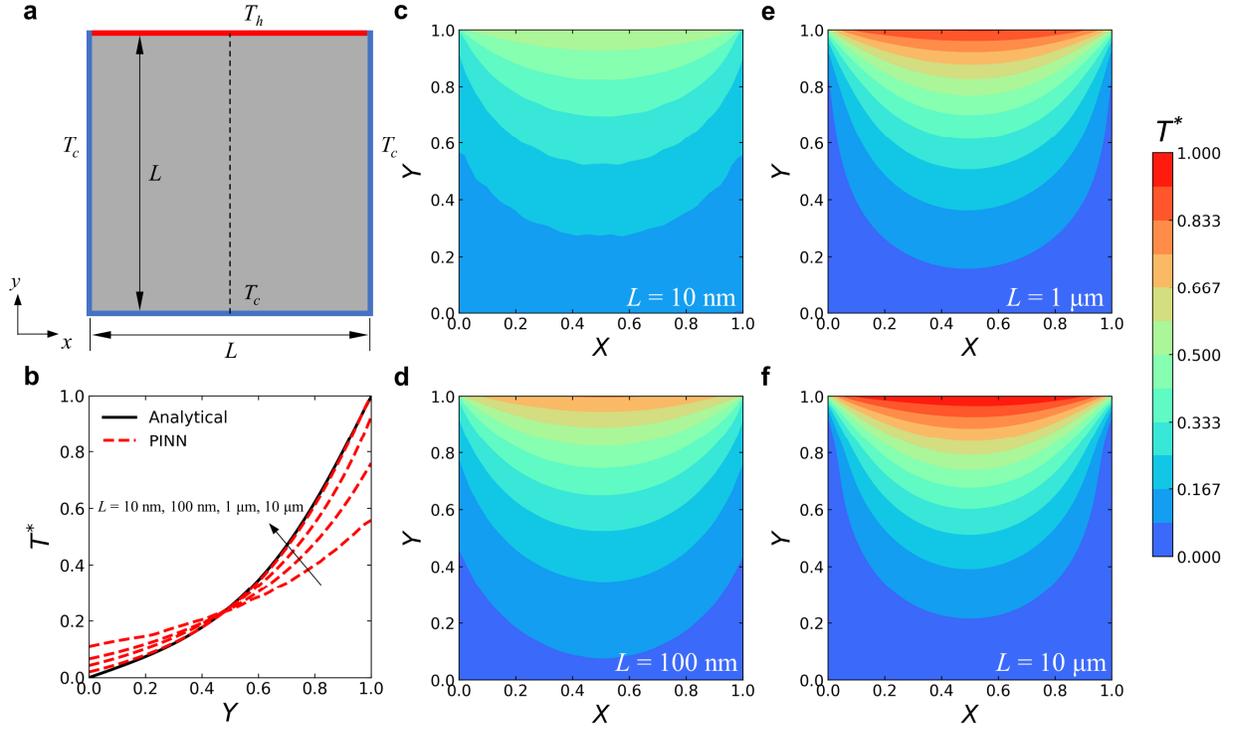

**Fig. 4 Results of 2D square phonon transport**. **a** Schematic of 2D square phonon transport case. **b** Non-dimensional temperature profiles along the vertical centerline (dashed line in **a**) at different length scales. Analytical solution is derived based on Fourier's law (i.e., solution in the diffusive limit). **c-f** Contours of dimensionless steady-state temperature at different length scales. *X* and *Y* are normalized spatial coordinates.

### 3D cuboid phonon transport

To assess the potential applicability of our PINN scheme to real devices, a 3D geometry is considered as depicted in Fig. 5a. The test geometry is a silicon block of dimensions $L \times L \times 0.5L$, with a circular hot spot of Gaussian temperature distribution on the top surface, while all the other surfaces are maintained at a lower temperature $T_c$. Such a hot spot may correspond to a hot spot in a semiconductor layer due to the Joule heating effect. The Gaussian temperature distribution has $T_{max}$ at the center of the top surface, and the temperature smoothly decreases to $T_c$ ($\Delta T = T_{max} - T_c = 1\ K$). The standard deviation of the distribution is set as $L/6$, with the full width at half maximum (FWHM) to be around 0.4$L$. Two PINNs are trained for the parametric learning of the multiscale thermal transport problem. One model is for the edge length $L$ between 10 nm and 1 μm, and the other is for $L$ between 10 μm and 1 mm. Each model is trained on three sample geometries of varying sizes. We note that separating the training for two smaller dimension regions can reduce the total number of training points and improve the prediction accuracy. 12 × 12 discrete angles are set for the training of the smaller geometries (3D cuboid[a] in Table 1), while 8 × 8 angles are used for the larger geometries (3D cuboid[b] in Table 1). This is because the phonon energy distribution is fairly isotropic in the diffusive regime, and we can use fewer angles without deteriorating the accuracy.

The steady-state temperature contour for the structure of length $L$ = 1 mm is shown in Fig. 5b, which is evaluated on a much denser grid mesh than the one used in training (see Table 1). It is observed that PINN successfully reproduces the Gaussian temperature distribution on the top surface (Fig. 5c). In Fig. 5d-i, we show temperature contours in the central planes (outlined by dashed lines in Fig. 5a) at different length scales. It is noted that 3D cases have rarely been investigated using BTE in previous studies, because it requires enormous computational resources. Thus, there is no solution available for comparison in this case. Nonetheless, we adopt a simple PINN model to obtain the solution of 3D heat conduction under the same boundary conditions based on the Fourier's law, as shown in Fig. 5i. This result can be treated as the benchmark solution in the diffusive limit since the final training loss is as low as $3\times10^{-4}$. Comparing PINN prediction in Fig. 5h with the benchmark in Fig. 5i, we find the error to be less than 1.4 % (Supplementary Fig. 2) as the 1 mm case should already be in the diffusive limit. Therefore, we expect high accuracies as well for geometries of smaller sizes. We note that it only takes a few minutes to solve the BTE using the PINN scheme on a single GPU for a wide length scale range. In comparison, it takes several hours to simulate a 3D system for a single time step on a parallel machine with 400 processors for a specific geometry in a previous numerical study[21].

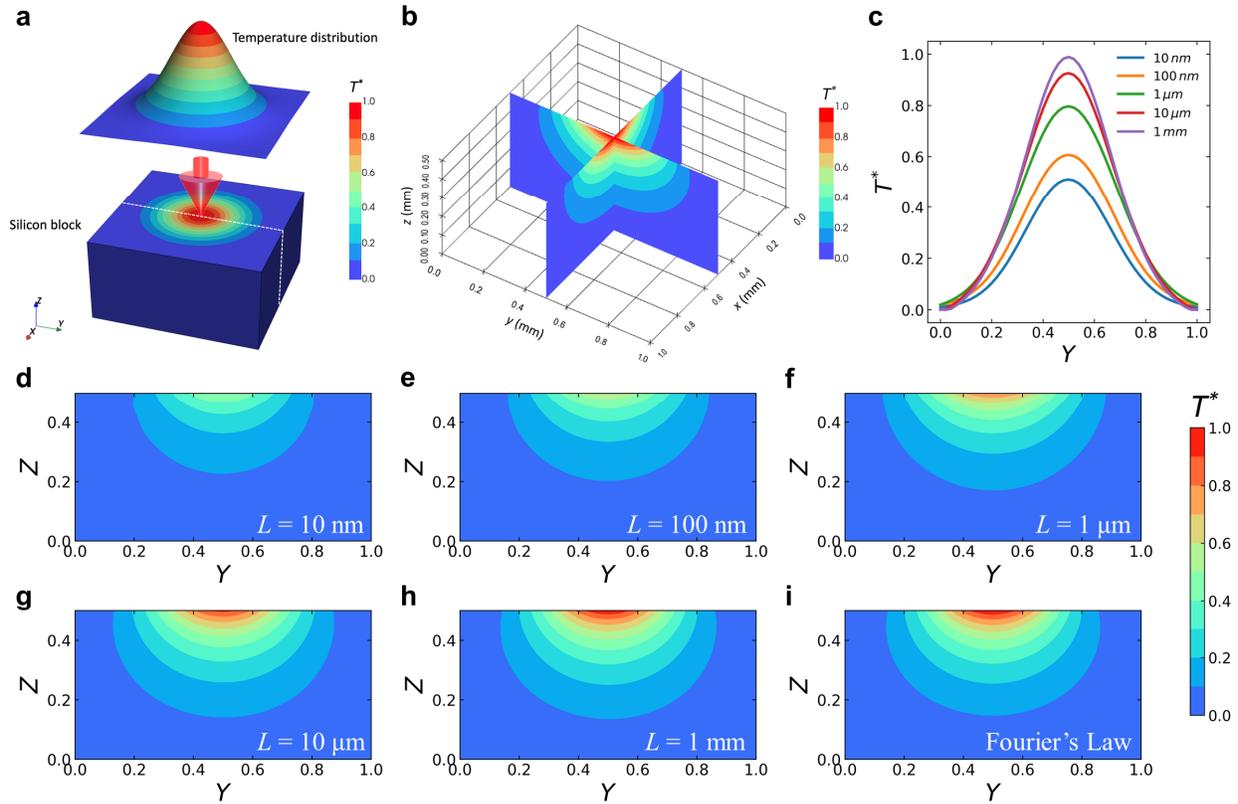

**Fig. 5. Results of 3D cuboid phonon transport**. **a** Schematic of the 3D phonon transport case. The system is a silicon block of dimensions $L \times L \times 0.5L$. Gaussian temperature distribution is applied to the top surface, while all the other surfaces are maintained at a lower temperature. **b** Dimensionless steady-state temperature distribution for a 3D system of size 1 mm × 1mm × 0.5 mm. **c** PINN-predicted dimensionless temperature distributions along the centerline on the top surface at different length scales. **d-h** PINN-predicted contours of dimensionless steady-state temperature in the central plane ($x = 0.5L$, outlined by dashed lines in **a**) at different length scales. **i** Solution of 3D heat equation based on Fourier's law under the same boundary conditions. The temperature distribution is obtained by a well-trained PINN. $Y$ and $Z$ are normalized spatial coordinates.

## 3. Discussion

As shown in Table 1, it can be seen that the training cost of PINN increases with the number of dimensions, but the online evaluation can be very fast even for high dimensional cases. We emphasize that this PINN scheme is a data-free ML method, circumventing the need for generation large amount of training data. Although conventional numerical methods can solve 1D and 2D problems accurately within a few minutes[31], they usually require long computation time (tens of hours) and large memory (hundreds of gigabytes) when dealing with 3D cases even using large scale parallel computing[21, 28]. In contrast to intricate parallel algorithms and computing, which may not be easily accessible, the proposed PINN model has the advantage of simple implementation and low evaluation cost. Moreover, by treating the characteristic length as an input parameter, this model is able to solve phonon BTE in a parametric setting, which will be beneficial for device design and optimization. We also note that since the phonon distributions are very different in the diffusive and ballistic regime, the prediction accuracy can be further improved by limiting the length scales of interest to a smaller range.

Although showing great promise, the current framework has several limitations, which warrants further research. First, this study is focused on parametrized steady-state phonon BTE, and the current model is not designed to handle dynamic systems. Further developments are necessary in order to solve spatiotemporal PDEs. For instance, we can employ the Long-Short Term Memory (LSTM) recurrent neural network architecture to predict the dynamics, whose effectiveness has been demonstrated in recent studies for time series prediction[64]. Second, the phonon BTE are solved under the assumption that the temperature difference is small enough such that the equilibrium state can be linearly approximated and the relaxation time is independent of space. To predict thermal transport with a large temperature difference, we need to adopt the distribution function *f*-based BTE with space-dependent relaxation time[30]. Third, the training cost will increase for complex problems (e.g., irregular geometries, multi-layered structures) since the equation residuals have to be calculated on massive amounts of collocation points in high-dimensional input spaces. We may adopt the convolutional neural network (CNN) structure[45, 65, 66] for complex geometries, which enables efficient learning even on irregular domains because of its parameter-sharing feature. While these directions will be explored in the future for practical applications, the current work represents the first demonstration of using PINN to efficiently solve mode-resolved phonon BTE. Our work may profoundly impact thermal transport research and make BTE a viable tool for practical device design and optimization.

## Methods

**Energy-based phonon BTE**

In general, the energy-based phonon BTE[4, 20, 29, 31, 67, 68], under the single mode relaxation time approximation, can be written as

$$\frac{\partial e}{\partial t} + \boldsymbol{v} \cdot \nabla e = \frac{e^{eq} - e}{\tau}, \tag{5}$$

where $e(\boldsymbol{x}, \boldsymbol{s}, k, p, t) = \hbar\omega D(\omega, p)[f - f^{BE}(T_{ref})]$ is the phonon energy deviational distribution function, $e^{eq}$ is the associated equilibrium distribution function, $\boldsymbol{v}$ is the group velocity, and $\tau$ is the effective relaxation time.

The phonon distribution function $f = f(\boldsymbol{x}, \boldsymbol{s}, k, p, t)$ (or $f(\boldsymbol{x}, \boldsymbol{s}, \omega, p, t)$) is a function dependent on the spatial vector $\boldsymbol{x}$, directional unit vector $\boldsymbol{s} = (cos\theta, sin\theta cos\varphi, sin\theta sin\varphi)$ ($\theta$ is the polar angle and $\varphi$ is the azimuthal angle), time *t*, wave number *k* (or angular frequency $\omega = \omega(k, p)$) and polarization *p*. $f^{BE}(T_{ref})$ denotes the equilibrium distribution function following the Bose-Einstein distribution at the reference temperature. Assuming that the temperature difference throughout the domain is much less than the reference temperature ($|\Delta T| \ll T_{ref}$), we can make the following approximation[20, 27]

$$e^{eq}(k,p,T) = \hbar\omega D(\omega,p)[f^{BE}(T) - f^{BE}(T_{ref})] \approx C(\omega,p)(T - T_{ref}),\tag{6}$$

where $\hbar$ is the Planck's constant divided by $2\pi$, $D(\omega,p) = \frac{k^2}{2\pi^2|v|}$ is the phonon density of states, and $C(\omega,p) = \hbar\omega D(\omega,p)\frac{\partial f^{BE}}{\partial T}$ represents the modal heat capacity. This assumption has been widely used in previous studies[29, 31, 68], and with that we can simplify the problem significantly by using the linear relation in Eq. (6). Furthermore, the relaxation time $\tau$ and the heat capacity $C$ can be treated as independent of space, and they are calculated based on the reference temperature $T_{ref}$.

In this work, we only consider the steady state. For a system without internal heat source, the first law of thermodynamics requires that the divergence of the heat flux must be zero, which can be calculated by integrating the right-hand side of Eq. (5) over the solid angle space ($\Omega$) and frequency space ($\omega, p$)

$$\nabla \cdot \boldsymbol{q} = \sum_p \int_0^{\omega_{max,p}} \int_{4\pi} \frac{e^{eq} - e}{\tau} d\Omega\, d\omega = 0,\tag{7}$$

where the total heat flux is

$$\boldsymbol{q} = \sum_p \int_0^{\omega_{max,p}} \int_{4\pi} \boldsymbol{v} e\, d\Omega\, d\omega,\tag{8}$$

and the local temperature can be obtained by substituting Eq. (6) into Eq. (7)

$$T = T_{ref} + \frac{1}{4\pi}\left(\sum_p \int_0^{\omega_{max,p}} \int_{4\pi} \frac{e}{\tau} d\Omega\, d\omega\right) \times \left(\sum_p \int_0^{\omega_{max,p}} \frac{C}{\tau} d\omega\right)^{-1}.\tag{9}$$

Due to non-equilibrium effects, the temperature calculated here is merely representative of the local internal energy instead of thermodynamics temperature[21, 29]. For the sake of brevity, it is still referred to as the temperature in numerical experiments. Based on Eqs. (5), (7) and (8) above, we can derive Eq. (1).

**Boundary conditions**

Generally, three types of boundary conditions are used in phonon transport problems[29, 69], including isothermal boundary condition, diffusely reflecting boundary condition and periodic boundary condition. These boundary conditions can be applied to problems with any parameter sets $\boldsymbol{\mu}$.

*Isothermal boundary* absorbs all phonons that strike it and emits phonons in thermal equilibrium with the boundary temperature $T_b$. Mathematically, this may be written as

$$e(\boldsymbol{x}_b, \boldsymbol{s}, k, p) = e^{eq}(k, p, T_b), \qquad \boldsymbol{s} \cdot \boldsymbol{n}_b > 0, \tag{10}$$

where $\boldsymbol{n}_b$ is the normal unit vector pointing into the simulation domain.

*Diffusely reflecting boundary* is a category of adiabatic boundary. At this boundary, the net heat flux is zero, and the phonons are reflected with equal probability along all possible directions, namely,

$$e(\boldsymbol{x}_b, \boldsymbol{s}, k, p) = \frac{1}{\pi} \int_{\boldsymbol{s}' \cdot \boldsymbol{n}_b < 0} e(\boldsymbol{x}_b, \boldsymbol{s}', k, p) |\boldsymbol{s}' \cdot \boldsymbol{n}_b| d\Omega, \qquad \boldsymbol{s} \cdot \boldsymbol{n}_b > 0. \tag{11}$$

When the *periodic boundary conditions* are applied, a phonon that crosses a periodic boundary is emitted at the opposite boundary with the same velocity vector and frequency. Moreover, the corresponding boundaries are subject to local thermal equilibrium, which can be expressed as

$$e(\boldsymbol{x}_{b_1}, \boldsymbol{s}, k, p) - e^{eq}(k, p, T_{b_1}) = e(\boldsymbol{x}_{b_2}, \boldsymbol{s}, k, p) - e^{eq}(k, p, T_{b_2}), \tag{12}$$

where $\boldsymbol{x}_{b_1}, T_{b_1}$ and $\boldsymbol{x}_{b_2}, T_{b_2}$ are the spatial coordinates and temperatures of two associated periodic boundaries $b_1$ and $b_2$, respectively.

**PINN architecture and training**

The proposed PINN model is composed of two sub-DNNs for the equilibrium ($e^{eq}$) and non-equilibrium ($e^{neq}$) parts of the phonon energy distribution, respectively (Fig. 1). The sub-DNNs share the identical structure of 8 hidden layers with 30 neurons per layer. Two sub-DNNs are trained simultaneously with a unified physics-informed loss function. We employ the Swish activation function $(x \cdot \text{Sigmoid}(x))$[70] in each layer except the last one, where a linear activation function is applied. The Adam optimizer[71], a robust variant of the stochastic gradient descent algorithm, is used to solve the optimization problem defined in Eq. (3) by randomly sampling mini-batches of inputs (see Supplementary Fig. 3 for an example of training process). The initial learning rate is set as $1 \times 10^{-3}$, and training points are generated by discretization of the input domain. To approximate the integrals in Eq. (1), the Gauss-Legendre quadrature[72] is adopted for the solid angle space, while the midpoint rule is used for the frequency space. In the case the spatial domain is logically rectangular, we can set the interior training points as low-discrepancy Sobol sequences[73] to alleviate the curse of dimensionality (Supplementary Note 2). Input spatial coordinates are scaled to the range [0, 1]. The output for the non-equilibrium part $e^{neq}$ is scaled with $Kn$ to make it of the order $O(1)$. The PINN algorithm is implemented within the PyTorch platform[74], and all numerical experiments are performed on a single NVIDIA GeForce TITAN Xp Graphic Processing Unit (GPU). The code and data will be available on GitHub upon publication.

**Acknowledgements**

The authors would like to thank ONR MURI (N00014-18-1-2429) for the financial support. The simulations are supported by the Notre Dame Center for Research Computing, and NSF through the eXtreme Science and Engineering Discovery Environment (XSEDE) computing resources provided by Texas Advanced Computing Center (TACC) Stampede II under grant number TG-CTS100078.

**Author contributions**

R.L., E.L. and T.L. conceived the idea and initiated this project. R.L. designed and trained the model. R.L., E.L. and T.L. discussed the results and wrote the manuscript.

**Competing interests**

The authors declare no competing interests.

**Data and code availability**

All data needed to evaluate the conclusions in the paper are present in the paper and/or the Supplementary Information. Data will be available in a GitHub repository upon publication. Additional data related to this paper may be requested from the authors.

11. Murthy JY, Narumanchi SV, Jose'A P-G, Wang T, Ni C, Mathur SR. Review of multiscale simulation in submicron heat transfer. *International Journal for Multiscale Computational Engineering* **3**, (2005).

12. Mazumder S, Majumdar A. Monte Carlo study of phonon transport in solid thin films including dispersion and polarization. *J Heat Transfer* **123**, 749-759 (2001).

13. Lacroix D, Joulain K, Lemonnier D. Monte Carlo transient phonon transport in silicon and germanium at nanoscales. *Physical Review B* **72**, 064305 (2005).

14. Mittal A, Mazumder S. Monte Carlo study of phonon heat conduction in silicon thin films including contributions of optical phonons. *Journal of Heat Transfer* **132**, (2010).

15. Péraud J-PM, Hadjiconstantinou NG. Efficient simulation of multidimensional phonon transport using energy-based variance-reduced Monte Carlo formulations. *Physical Review B* **84**, 205331 (2011).

16. Escobar RA, Amon CH. Influence of phonon dispersion on transient thermal response of silicon-on-insulator transistors under self-heating conditions. (2007).

17. Escobar RA, Amon CH. Thin film phonon heat conduction by the dispersion lattice Boltzmann method. *Journal of Heat Transfer* **130**, (2008).

18. Murthy J, Mathur S. Computation of sub-micron thermal transport using an unstructured finite volume method. *J Heat Transfer* **124**, 1176-1181 (2002).

19. Narumanchi SV, Murthy JY, Amon CH. Submicron heat transport model in silicon accounting for phonon dispersion and polarization. *J Heat Transfer* **126**, 946-955 (2004).

20. Minnich AJ, Chen G, Mansoor S, Yilbas B. Quasiballistic heat transfer studied using the frequency-dependent Boltzmann transport equation. *Physical Review B* **84**, 235207 (2011).

21. Ali SA, Kollu G, Mazumder S, Sadayappan P, Mittal A. Large-scale parallel computation of the phonon Boltzmann Transport Equation. *International journal of thermal sciences* **86**, 341-351 (2014).

22. Péraud J-PM, Hadjiconstantinou NG. Adjoint-based deviational Monte Carlo methods for phonon transport calculations. *Physical Review B* **91**, 235321 (2015).

23. Chattopadhyay A, Pattamatta A. A comparative study of submicron phonon transport using the Boltzmann transport equation and the lattice Boltzmann method. *Numerical Heat Transfer, Part B: Fundamentals* **66**, 360-379 (2014).

24. Guo Y, Wang M. Lattice Boltzmann modeling of phonon transport. *Journal of Computational Physics* **315**, 1-15 (2016).

25. Adams ML, Larsen EW. Fast iterative methods for discrete-ordinates particle transport calculations. *Progress in nuclear energy* **40**, 3-159 (2002).

26. Loy JM, Mathur SR, Murthy JY. A coupled ordinates method for convergence acceleration of the phonon Boltzmann transport equation. *Journal of Heat Transfer* **137**, (2015).

27. Loy JM, Murthy JY, Singh D. A fast hybrid Fourier–Boltzmann transport equation solver for nongray phonon transport. *Journal of heat transfer* **135**, (2013).
26

Supplementary Information for

**Physics-Informed Neural Networks for Solving Multiscale Mode-Resolved Phonon Boltzmann Transport Equation**

Ruiyang Li, Eungkyu Lee*, and Tengfei Luo*

* Corresponding author. Email: eleest@khu.ac.kr (E. Lee); tluo@nd.edu (T. Luo)

**Supplementary Note 1: Phonon dispersion and scattering**

The dispersion relations of the acoustic phonons are approximated as[1]

$$\omega = c_1 k + c_2 k^2,$$

where for LA branch $c_1$ = 9.01 × 10$^5$ cm/s, $c_2$ = −2.0 × 10$^{-3}$ cm$^2$/s; for TA branch $c_1$ = 5.23 × 10$^5$ cm/s, $c_2$ = −2.26 × 10$^{-3}$ cm$^2$/s. The Matthiessen's rule is used to estimate the effective relaxation time by combining different scattering processes[2], including the impurity scattering, umklapp (U) and normal (N) phonon-phonon scattering,

$$\tau^{-1} = \tau_{impurity}^{-1} + \tau_U^{-1} + \tau_N^{-1} = \tau_{impurity}^{-1} + \tau_{NU}^{-1},$$

where the relaxation time formulas and coefficients are given in Supplementary Table 1. We note that the dispersion and relaxation times can also be from first-principles calculations for each discrete mode[3]. The PINN implementation will not change.

**Supplementary Table 1 | Relaxation time formulas and coefficients[4].**

| $\tau_{impurity}^{-1}$ | $A_i \omega^4$, $A_i = 1.498 \times 10^{-45}$ s$^3$ |
|---|---|
| LA | $\tau_{NU}^{-1} = B_L \omega^2 T^3$, $B_L = 1.180 \times 10^{-24}$ K$^{-3}$ |
| TA | $\tau_{NU}^{-1} = B_T \omega T^4$, $0 \leq k < \pi/a$ |
| | $\tau_{NU}^{-1} = B_U \omega^2 / \sinh(\hbar\omega/k_B T)$, $\pi/a \leq k \leq 2\pi/a$ |
| | $B_T = 8.708 \times 10^{-13}$ K$^{-3}$, $B_U = 2.890 \times 10^{-18}$ s |
| | $a = 0.5431$ nm |

**Supplementary Note 2: Training points in the spatial domain**

We use low-discrepancy Sobol sequences[5] for logically rectangular spatial domains to constitute the interior training sets. Sobol sequences arise in the context of Quasi-Monte Carlo integration[6], which can alleviate the curse of dimensionality and provide more flexibility for high-dimensional problems. Figure S1 shows the spatial domains containing interior training points in 2D phonon transport problems. In 2D in-plane cases, we use 300 Sobol points for the interior domain, and 30 equidistant points for each boundary. In 2D square cases, for better description of phonon transport, we construct a non-uniform spatial domain (450 Sobol points) with more training points close to the hot boundary.

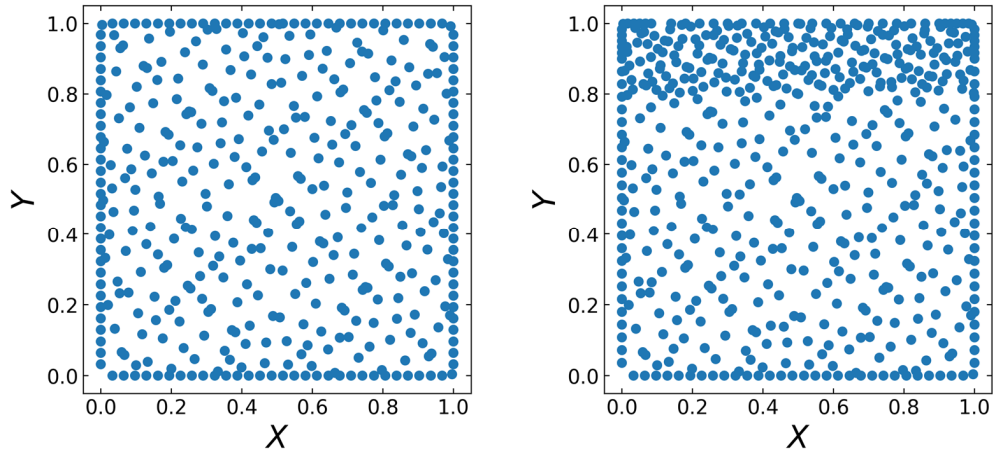

**Supplementary Figure 1** | Spatial training points based on Sobol sequences for **(left)** the 2D in-plane case, and **(right)** the 2D square case.

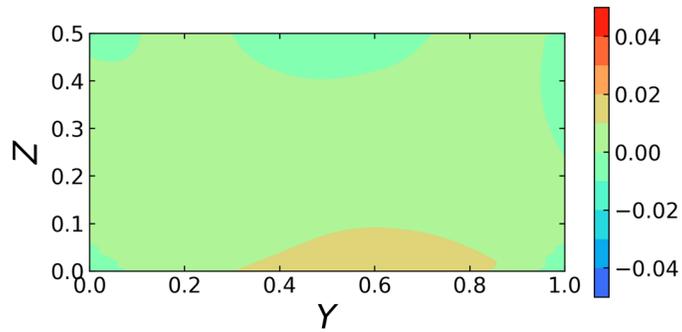

**Supplementary Figure 2** | Difference between PINN predicted temperature contour (in a 3D system of size 1 mm × 1mm × 0.5 mm) and solution of 3D heat equation based on Fourier's law in the central plane. The error rate is less than 1.4 %.

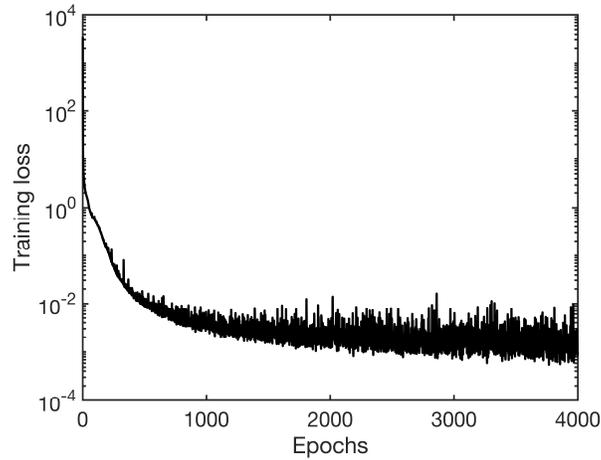

**Supplementary Figure 3** | Example of learning curve in 2D in-plane phonon transport problem. The training loss is averaged among 15 mini-batches for each epoch.